# Synergy from reproductive division of labor and genetic complexity drive the evolution of sex

Klaus Jaffe

Universidad Simón Bolivar, Caracas

kjaffe@usb.ve

# Abstract

Computer experiments that mirror the evolutionary dynamics of sexual and asexual organisms as they occur in nature, tested features proposed to explain the evolution of sexual recombination. Results show that this evolution is better described as a network of interactions between possible sexual forms, including diploidy, thelytoky, facultative sex, assortation, bisexuality, and division of labor between the sexes, rather than a simple transition from parthenogenesis to sexual recombination. Diploidy was shown to be fundamental for the evolution of sex; bisexual reproduction emerged only among anisogamic diploids with a synergistic division of reproductive labor; and facultative sex was more likely to evolve among haploids practicing assortative mating. Looking at the evolution of sex as a complex system through individual based simulations, explains better the diversity of sexual strategies known to exist in nature, compared to classical analytical models.

# Author summary

Computer experiments that mirror the evolutionary dynamics of sexual and asexual organisms showed:

1- Evolution is better described as a network of interactions between possible sexual forms, including diploidy, thelytoky, facultative sex, assortation, bisexuality, and division of labor between the sexes, rather than a simple transition from parthenogenesis to sexual recombination.

2- Diploidy was shown to be fundamental for the evolution of sex

3- Bisexual reproduction emerged only among anisogamic diploids with a synergistic division of reproductive labor

4- Facultative sex was more likely to evolve among haploids practicing assortative mating. Looking at the evolution of sex as a complex system through individual based simulations explains better the diversity of sexual strategies known to exist in nature, compared to classical analytical models.

# Introduction

The emergence of sex is considered one of the major transitions in evolution [1], but the adaptive value of sex is still a mystery. Analytical theoretical biology has struggled with this issue for a long time [2-3], but our understanding of the evolution of sexual recombination is still very partial and incomplete. Many models mostly based upon very oversimplified and unrealistic parameters have been published. They served to define several important concepts that now seem to have been broadly accepted. The Red Queen hypothesis or constant adaptation to survive against ever-evolving opposing organisms [4], has been popular but is not sufficient to explain the ubiquity of sex [5]. The most important hypothesis is that sex uncouples beneficial from deleterious mutations, allowing selection to proceed more effectively with sex than without it [6]. A new revision of empirical evidence on sex handling deleterious mutations successfully, corroborates this view [7]. The analysis of synergistic epistasis has been important in evolutionary genetics, but has been focused mainly on interactions between deleterious mutations in different gene loci [8-12]. Several complex issues remain to be resolved [13]. For example, the effect of synergy emerging from the interactions between the sexes, is poorly understood [14-15]. Synergy is defined by the Oxford dictionary as "The interaction or cooperation of two or more organizations, substances, or other agents to produce a combined effect greater than the sum of their separate effects". Here we will explore the effect of synergistic interactions among the sexes and try to understand the difference in the evolution of haploids (the most common assumption in the literature) versus that of diploids (the most common form found in nature) on this evolutionary dynamic.

The simpler an explanation, the better. Sometimes however, excess simplicity eliminates the elements needed to understand a phenomenon. It has been argued for a long time that analytical tools that proved successful for analyzing problems with few variables are not appropriate for the study of complexity [16]. Computational methods may overcome some of these problems in biology and elsewhere [17]. Sex is a complex adaptive strategy that allows evolution to navigate rough fitness landscapes by optimizing recombination to produce offspring with increased fitness. Tools promoted by complex system sciences, such as computer experiments and Agent Based Modeling (ABM) have been successful in allowing new insights into these problems by showing, for example: that selection in the presence of sex favors the maintenance of synergistic interaction between genes in a highly robust manner. For example, Livnat et al. [18] studied "*the ability of alleles to perform well across different combinations*". This ability can be viewed as a kind of synergy in positive epistasis (the contrary of synergy among deleterious mutations). ABM also showed the importance of multi-level sexual selection, both above the individual level [19] and below the individual level such as in gamete selection [20]; and the importance of assortative mating [14] in maintaining the working of epistatic genes (where the effect of one gene depends on the presence of one or more 'modifier genes').

Assortation, as an element of inclusive fitness that includes assortative mating [21], is more general than kin-selection and includes kin selection [22]. Assortation allows sex to select synergistic combinations of alleles, increasing the "Error Thresholds" or critical mutation rate beyond which structures obtained by an evolutionary process are destroyed more frequently than selection can reproduce them [23]. This phenomenon has also been called homophily, assortation, narcissism and "similarity selection" and has important effects on the evolution of sex [24].

The synergy through behavioral and genetic cooperation between the sexes, modulated by anisogamy, is important in understanding biological and economic processes [22, 25]. This study focuses on the effect on fitness of two sexual partners that is greater than expected from their effects when alone. This effect is referred as synergy here. Thus, I analyze here the emergence and evolution of sex with computational experiments that work analogously to a supercollider of ideas [26], where different hypotheses for the evolution of sex are tested against each other.

# Methods

For simulations, I used Biodynamica, a robust metaphor for biological evolution. An older version of this program mirrored successfully the different optimal strategy of biocide application to retard the emergence of resistance to biocides in asexual viruses and sexual insects [27]. The model creates populations of agents or virtual organisms, each one possessing a genome with different genes. Each gene had an allele coding for a specific behavior or other phenotypic characteristic (Table 1). A gene coding for the type of sexual strategy the agent used (gene 1 in Table 1) could be occupied by one of five different alleles coding for either: asexual reproduction by cloning; monosexual reproduction by thelytoky;

bisexual reproduction, as among most living organisms, including gametogenesis and mitotic recombination; haplo-diploidy where females were diploid and males haploid as in the Hymenoptera; and "hermaphrodites" practicing facultative sex, so that they out-crossed with male or with another hermaphrodite.  Gene 2 coded for ploidy (number of sets of chromosomes in the genome), with alleles for either haploidy, diploidy or haplodiploidy. Simulation for sexual diploids reproduction included explicit simulation of gametogenesis, mitosis, meiosis, and crossovers between parent's gametes during fertilization. The phenotype coding of alleles in the other simulated genes are listed in Table 1. Thus, simulations mirrored as closely as possible the mechanisms of sexual recombination known to occur in nature, including gametogenesis, mitosis, random crossovers, and mutations. A simplified pseudo-code is given in Table 2.

Genes 1 to 7 defined a population of agents that are susceptible to being killed by 3 different biocides and with varying forms of ploidy and sexual strategy. Gene 8 to 11 defined phenotypes that determined characteristics of the life history of agents. Genes 12 to 15 determined mate selection and parental investment behavior. Specifically Gene 15 codes for "bisexual social synergy" which is the synergy unleashed by social interactions between sexes;  and/or "synergistic anisogamy" which is the synergy unleashed by anisogamy.

Phenotype expression was based on the alleles in the single chromosome of the genome in haploids. In diploids, only a single randomly selected allele for each loci was expressed phenotypically. Simulations consisted in letting 600 agents mate or clone according to the different rules coded in their alleles, reproduce, suffer random death, death from biocides, deadly mutations, and lethal combination of alleles. Experiments consisted in creating an initial population of agents with a homogeneous frequency distribution of alleles in a specific set of genes. Selection and reproduction at each time step varied this frequency distribution. The program allowed us to observe the evolution of the allelic composition of the population during a period. The most successful combination of alleles were the ones that reproduced more and survived selection better at every time step. The higher the population size, the larger the number of random deaths, so that populations maintained a size of around 600 individuals.

**Table 1**: Gene loci of the genome of agents, and the possible alleles for each them

| Loci | Gene | Phenotypic expression of allele according to Number | Range of variance |
|---|---|---|---|
| 1a | Sexual Strategy | 0: Asexuals<br>1: Monosexuals<br>2: Bisexuals<br>3: Sexual-Asexual (as in haplodiploidy): females are produced sexually and males asexually<br>4: Sexual hermaphrodites (hermaphrodites mate only with other hermaphrodites)<br>5: Sexual hermaphrodites (hermaphrodites mate only males and with other hermaphrodites) | 0-5 |
| 2a | Ploidy | 1: Haploid<br>2: Diploid<br>3: Triploid. Are explored in [28] | 1-2 ** |
| 3 | Sex | 1: female<br>2: male | 1-2 |
| 4 | Mutation probability | Mutation rates at probabilities according to formula: $p = 0.2 \wedge (allele +1)$ | 0-2 |
| 5 | Resistance 1 | Resistance is given in a continuous range so that allele 0 is the most resistant (i.e. is immune) and allele 10 is the least resistant to biocide 1. Concentration of biocide fluctuates randomly | 0-10 |
| 6 | Resistance 2 | Only allele 0 is resistant and all other alleles are susceptible to biocide 2. Concentration of biocide fluctuates randomly | 0-10 |
| 7 | Resistance 3 | Only allele 0 is resistant and all other alleles are susceptible to biocide 3. Concentration of biocide fluctuates randomly | 0-10 |
| 8b | Life Span | Number gives the time steps of the maximum possible life expectancy of the individual. | 0-10 or 10 * |
| 9b | Clutch size | Number of offspring at each reproductive act. | 0-10 or 10 * |
| 10b | Reproductive age Female | Nr of time steps after which reproduction starts for females | 0-5 or 1 * |
| 11b | Reproductive age Male | Nr of time steps after which reproduction starts for males | 0-5 or 1 * |
| 12c | Mating Effort | Number of males (or females in hermaphrodites), screened for mating according to criteria defined by gene 13. MV = 0 or = 1 will screen just 1 individual. | 1-100 ** |
| 13c | Mate Selection Criteria | 0: Random selection of mates. Female mates with the first male encountered<br>1: Female mates only with males with the same Sexual Strategy allele (gene loci 1). Females prefer males with good resistance genes and mate assortatively regarding the other genes<br>2: Open assortment as in 1, but females mated with males with any Sexual Strategy. Sexual strategy of female was inherited to offsring. | 0-1 ** |
| 14 | Amount | Amount of fitness increase provided to its offspring | 0-2 ** |

| | Parental Investment | Increase of offspring fitness = Allele Nr /10 | |
|---|---|---|---|
| 15 | Bisexual Social Synergy | 0: No social synergy<br>1: Doubles the fitness of bisexual offspring as a metaphor of synergistic anisogamy without cost to the parent. | 0-1 ** |

a, b, c indicate that genes with the same letter are in the same epistatic group

* indicates that allele was fixed at this value in Simple Experiments

** indicates the variance in a range as used in Table 3

**Table 2: Simplified Pseudo-code** (for the compete code see note at the end of Methods)

1. **Initiation**: Random assignment of alleles to genes of individuals in the initial population
2. **Selection**: Individuals were excluded from the population when any of the following criteria was true:
    1. Their age exceeded their genetically prefixed life span.
    2. They were randomly selected by density independent criteria. For example I out of each 100 individuals chosen at random was killed, and this rate was increased logaritmically with increasing density.
    3. The density dependent selection criteria where tuned so as to deviate no more than 5% of the initially fixed optimal population size. The phenotype of the organism affected survival probability. For example, better nurtured offspring had higher probabilities of survival.
    4. Individuals not expressing the resistant allele of gene R1, R2, and R3 were killed with a probability pe1, pe2, and p3 which varied randomly each time step between 0 and 0.9.
3. **Mating**: All females select one mate according to their alleles in loci 12 and 13. Thus, mating was assortative (like with like) or at random. Asexuals reproduced without mating. Mating was between the same mating types except when allele 13-2 was present.
4. **Reproduction**: Females reproduce according to their ploidy and sexual strategy
5. **Cost of sex**: Low mating effort, determined by alleles in loci 12, increased failed reproduction with no offspring, as no appropriate mating partner was found. Asexuals had no males and always produced offspring. Thus, a given number of asexuals produced at least twice the number of offspring than the same number of sexuals.
6. **Variation**: New born individuals suffered random mutations at randomly chosen genes
7. **Recurrence**: Go to step 2 until maximum time steps have been achieved

The simulations track the evolutionary process at the level of genes. Each simulation creates a population of agents or organisms with different phenotypes in accordance to their allelic composition and aggregates the data at the population level. Each simulation was run with random initial conditions, where alleles were distributed uniformly randomly in each locus, according to the ranges given in Table 3. The outcome in most cases was that a specific sexual strategy eventually dominated the allele pool completely. Dominant sexual strategy made themselves evident after approximately 200 time steps. Therefore, the averages of the frequency of alleles among 100 repeated simulations during 400 time steps were shown (Figure 1). The standard deviation of the average was normally less than 30 % of the mean.

More simulations with conditions selected at will can be run by the reader. By using either the Unity program or the one written in VB6. The Unity version of Biodynamica can be downloaded or used directly online at http://bcv.cee.usb.ve/juegos/biodyn_en.html. The compiled Visual Basic version of Biodynamica used for the quantitative experiments reported, here can be downloaded for use in a Windows environment at http://atta.labb.usb.ve/Klaus/Programas.htm, together with the VB6 code.

# Results

The simulation results show that the fate of alleles coding for a sexual strategy is very susceptible to the possible range of allelic composition of agents in the population. The complexity of the simulated genome, quantified in number of loci, strongly affected the equilibrium frequency distribution of alleles (Table 3). In the populations composed of agents with the simplest genome (Exp 0), haplo-diploid sexual strategies were the most successful. Increasing complexity of the genome but maintaining all other conditions the same (Exp 1s) made asexuals the most successful. In populations composed of agents with an even more complex genome (Exp 1C) asexuals dominated strongly (see Fig 1). Table 3 shows experiments 1 to 4 in both the simple and the complex genome version. Clearly, complexity favored the likelihood for asexual to dominate in all cases. Experiment 5 tested the evolution of populations composed exclusively of haploids. Here, the level of genetic complexity seemed to be less relevant in the resulting sexual strategy favored by selection (see Exp 5S and 5C in Table 4). In experiments 6 to 10, the impact on evolution of alleles that affect mating behavior and parental investment were tested. The results can be seen in Table 3 and in Figure 1.

**Table 3:** The simulated genes and their allelic variance. Each allelic value coded for a specific phenotype. All cases involve genes 2, 12, 13, 14 and 15. In experiment 0 individuals also have genes 1 to 5. In experiment 1 they have 1 to 7 and in C 1 to 12. For example, allele 1 of gene 2 coded for haploid agents, whereas allele 2 coded for diploid ones. The experiment number correspond to the one in Figure 1.

| | | Allowed range of values for alleles | | | | | | | | | | | | | | |
|---|---|---|---|---|---|---|---|---|---|---|---|---|---|---|---|---|
| | | Loci 1-5 | Simple genome (loci 1-7) + | | | | Complex genome (loci 1-12) + | | | | | | | | | |
| Gene | Experiment | 0 | 1S | 2S | 3S | 4S | 5S | 1C | 2C | 3C | 4C | 5C | 6C | 7C | 8C | 9C | 10C |
| 2 | Ploidy | 1-2 | 1-2 | 1-2 | 1-2 | 1-2 | 1-1 | 1-2 | 1-2 | 1-2 | 1-2 | 1-1 | 1-2 | 1-2 | 1-2 | 1-2 | 1-2 |
| 12-13 | Mate Selection | 0-0 | 0-0 | 0-0 | 0-0 | 0-1 | 0-1 | 0-0 | 0-0 | 0-0 | 0-1 | 0-1 | 0-1 | 0-1 | 0-0 | 0-1 | 0-1 open |
| 14 | Parental Investment | 0-0 | 0-0 | 0-0 | 0-2 | 0-0 | 0-2 | 0-0 | 0-0 | 0-2 | 0-0 | 0-2 | 0-2 | 0-0 | 0-2 | 0-2 | 0-2 |
| 15 | Synergy | 0-0 | 0-0 | 0-1 | 0-0 | 0-0 | 0-1 | 0-0 | 0-1 | 0-0 | 0-0 | 0-1 | 0-0 | 0-1 | 0-1 | 0-1 | 0-1 |
| Dominant | | HDip | Asex | Bisex | HDip | Asex | Herm | Asex | Asex | Asex | Asex | Herm | Herm | Bisex | Asex | Bisex | Bisex |
| Sub-dominant | | Bisex | HDip | HDip | Bisex | HDip | Monos | Monos | Bisex | Monos | Monos | Monos | HDip | Asex | Bisex | Herm | Asex |

**Dominant**: Alleles are present in average in more than 60 % of genomes of the population. Average of 100 simulations

**Sub-dominant**: The given alleles are frequent but present in average in less than 50% of the genomes in the population

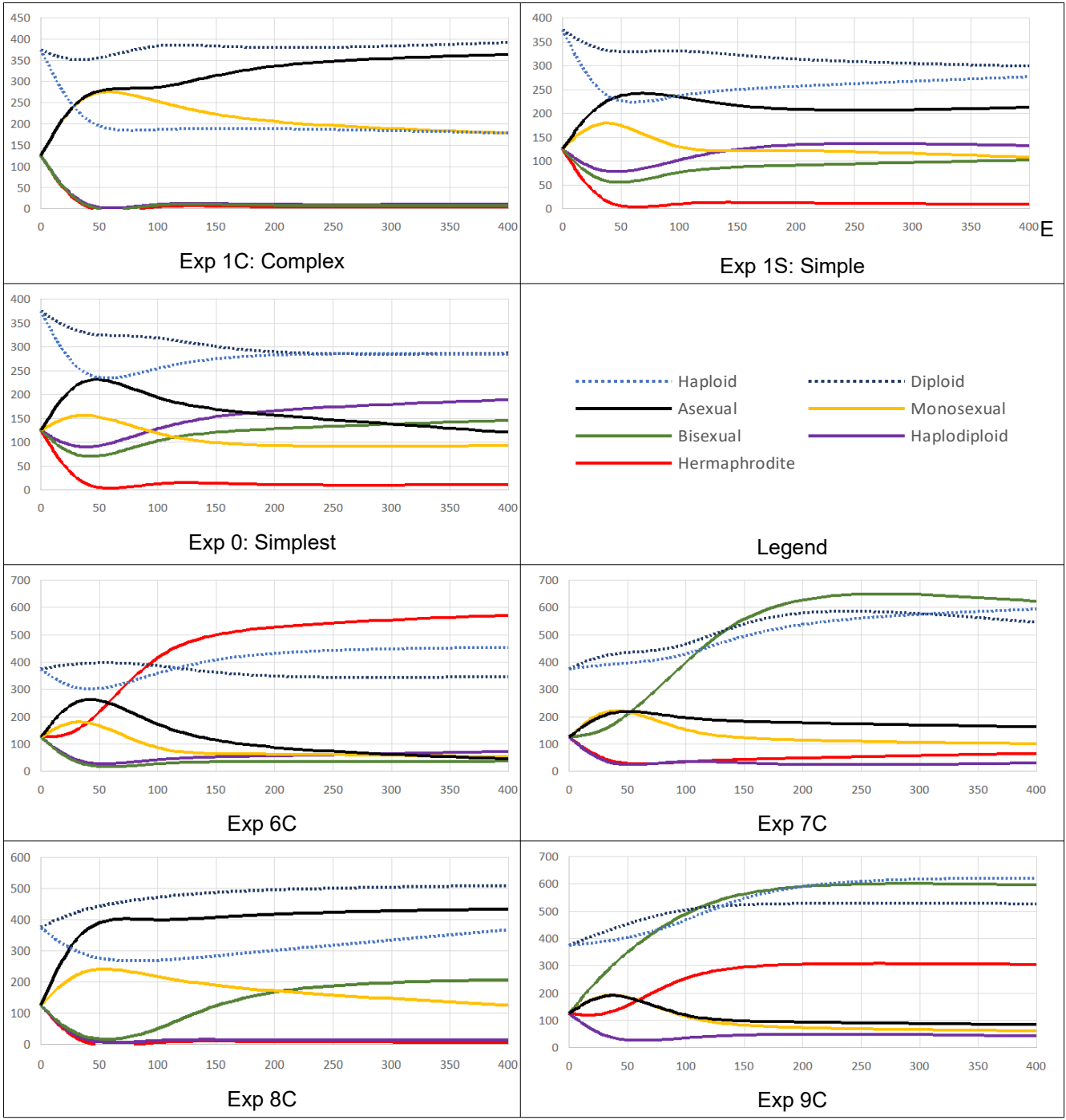

**Figure 1:** Curves show the average of 100 simulations of the number of copies of a given allele in the population of 600 agents in different computer experiments as given in Table 3. All simulations started with a random homogeneous proportion of all alleles and were run for 400 time steps. The x axis indicate the number of time steps. The y axis gives the number of copies of a given allele for sexual strategy and for ploidy.

Table 4 presents a statistical summary of results from all 15 experiments. Here, for each sexual strategy, the probability that the allele coding for it became dominant, i.e. became by far the most frequent in the population, is given. The Pearson correlation coefficient relates the likelihood for a given sexual strategy to become dominant with the presence of other genes. The second column, for example, shows that alleles coding for Asexual, Sexual, and Haplo-Diploids became dominant only among diploids (with 100% probability), whereas Hermaphrodites prospered only among haploids (with 70% probability). Column 5 shows the results of simulations when genes for synergy were also present: bisexuals became dominant with a probability of 100% (Pearson correlation = 1); whereas Haplo-Dipolid alleles never became dominant when genes for synergy co-occurred (Pearson correlation = 0).

**Table 4:** Summary of results from Table 3. Probability of co-occurrence of a given allele in 4 different loci, with the dominant sexual strategy as calculated from data shown in Table 3. P of 1 indicate 100% occurrence, whereas 0 indicates that this was never observed.

| Dominant Sexual Strategy | Gene2: Allele for Diploidy * | Gene 12: Allele for Mate Selection | Gene 14: Allele for Parental Investment | Gene 15: Allele for Sex Synergy | Complexity: Variance in alleles of gene loci 1-12 |
|---|---|---|---|---|---|
| Asexual | **1.0** | 0.3 | 0.3 | 0.3 | 0.7 |
| Bisexual | **1.0** | 0.7 | 0.3 | **1.0** | 0.7 |
| Haplo-Diploid | **1.0** | **0.0** | **1.0** | **0.0** | **0.0** |
| Hermaphrodites | 0.3 | **1.0** | **1.0** | 0.7 | 0.7 |

* No diploidy allele means allele for haploidy were simulated.

The table shows a series of interesting correlations:
1- Bisexuals became dominant in experiments that allowed the simultaneous presence of alleles for synergy and for diploidy in the genome of the agents.
2- Hermaphrodites (facultative sex) dominate the evolutionary outcome in populations with alleles for mate selection and for parental investment.
3- Haplo-diplods dominated the evolutionary process in populations of agents with simple genomes, parental investment and when synergy between sexes was absent.
4- In general, the evolution of sexual strategies was very dependent on the ploidy of the simulated genome.

Alleles for parental investment and mate selection have a delayed effect. It is the offspring who increases the odds of survival from the presence of the allele, not the parent. The presence of this effect influenced the success of sexual strategies. These results support the conjecture that asexual reproduction is better for short term selection for survival, whereas sexual reproductions is better for accumulating genes that have a delayed effect on fitness. This might explain why asexual reproduction is more successful than the sexual kind in populations of agents with complex genomes that lack alleles with delayed effects as reported above.

The results clearly show that a synergistic division of labor between the sexes favor alleles coding for bisexuality among diploids but not among haploids. Here, offspring of bisexual parents have an increased fitness due to parents offering parental investment. If this proxy for a synergistic division of reproductive labor is absent, and if parental investment is allowed, facultative sex (hermaphrodites) displaces bisexuality as the most successful sexual strategy.

Evolution of sex is affected by sexual selection [29], mate selection [30], and specifically assortative mating [31]. Results showed that assortation or homophily, and mate and sexual selection, strongly favored the evolutionary establishment of sex.

In experiments 2-9, females selected mates that shared their type of sexual strategy. Eliminating this restriction and allowing females to mate with males with different sexual strategies (Exp 10) increased the likelihood for sex to become the dominant strategy (Table 3 and supplementary material). This is due to a reduced cost in finding mates and thus, failing to reproduce, in experiment 10.

## Discussion

Many studied on the evolution of sex have been published. To cover them, I cited only the most extensive review [2], and the most recent one [7]. Despite this abundance of studies, few models, deal with diploid organisms [32-39]. This reflects the difficulty of tackling analytically the evolution of diploids with complex genomes possessing more than 3 loci. Numerical computer calculation, however, can tackle these problems. The results of such calculations shown here, is that without diploidy, sex is less likely to emerge through evolution. One reason for this is that diploidy mitigates the reported reduction of genetic variation by sex [40]. In the simulations presented, diploidy reduced the impact of selection on a given allele, prolonging its survival, and thus increasing the chance for possible synergistic interactions between different alleles to appear

Many adaptation rather than a single factor, including diploidy, thelytoky, facultative sex, assortation, bisexuality, and division of labor, explains better the emergence of the diversity of sexual strategies that exist in nature. The simulation results showed that although asexuality speeds adaptation of viable genotypes in complex settings, optimal conservation of genotypes with synergistically interacting alleles is favored by sex. The balance between these two forces may determine the specific evolutionary route to sexual reproduction taken in each environment.

The most relevant novel finding, in addition to the importance of diploidy, is that without the synergy unleashed between sexual partners, providing a better combination of genes to their offspring and making parental investment more efficient, bisexuality would not be superior to facultative sex in adapting to complex changing rough fitness landscapes. The chance for the emergence of synergy is enhanced by a greater store of diverse alleles achieved with diploidy.

The role of synergy is ubiquitous in biology and economics. Social Synergy accelerates evolution [41-42] and is the basis of biological and economic dynamics [22]. The production of synergy requires division of labor, including division of reproductive labor. This is the importance of anisogamy in evolution [43]. Male gametes optimize movements to find female gametes, which in turn optimize accumulation of resources, such as yolk. Anisogamy also refers to any sex-specific specialization in anatomy or behavior that increase the efficiency in the cooperation between sexes, leading to a fitness increase of their offspring. The logic behind this assumption is that both sex specific tasks cannot feasibly be performed simultaneously, and synergy arises through Adam Smith's invisible hand produced by division of labor [44]. Increasing evidence shows divergent adaptive pressures among the sexes [24]. Other benefits from this division of labor have been proposed. For example, Atmar [45], showed that cheap-to-produce males in sexual populations could be used to weed out deleterious mutations. A preliminary review of the occurrence of parental investment in nature seems to corroborate that bisexual species are more likely to show parental investment than asexual ones, and that haploids are less likely to be bisexual than diploids, but a rigorous systematic review is in order.

Among the reasons for the effect of diplody on the evolutionary dynamics of sex, is that sexual diploids have twice as many loci for hosting alleles than asexuals. Among haploids, alleles that have long term effect such as those regulating parental investment disappear before they can show their usefulness because selection focuses first on allelic combinations that guarantee immediate survival (resistance to biocides or large clutch sizes in the present model). Diploids have more loci to conserve alleles that might be useful in the future. This difference is more striking when considering the search work an evolutionary simulated process is required to perform in relation to the size of the allelic combinatorial landscape needed to explore. The simple genome with 7 loci allowed for $8.2 \times 10^5$ unique allelic combination, whereas the full complex version with 15 loci allowed $1.6 \times 10^9$ unique allelic combinations. Each individual diploid can test in each generation up to two times more alleles to explore these landscapes than haploids. This difference is compounded for each successive generation. The results showed that this advantage was more noticeable in more complex environments. That is, diploid sexual strategies increase the likelihood of finding an optimal combination of alleles in large allelic combinatorial landscapes, whereas haploid asexual strategies are more efficient in finding fast sub-optimal but effective combinations that assure survival. Poliploidy though has a limit: excess allelic redundancy hinders adaption as simulations with triploids showed [28]. Empirical evidence supporting this finding comes from organisms that can switch from asexual to sexual reproductive strategies. They favor asexual reproduction over the sexual kind when the adaptive pressures they suffer become more challenging [46-47].

      For the understanding of evolution in general, sexual recombination is fundamental. The emergence of sex together with assortative mating might have had a role in milestones of evolutionary history [48], such as the Cambrian explosion [49]. The high diversity of sex determination systems [50] is proof that sex has evolved through different pathways driven by different factors. The computer experiments presented here are compatible with this view of several feed-back loops, conforming a network of factors that affect the adaptive value of sexual strategies. Understanding the working of sexual recombination in its multiple forms has important practical applications, such as controlling malaria vectors [51], managing resistance to pests' pheromones [52] or biocides [27], or understanding the presence of "kings" and "queens" among social insects [53].

# Conclusions

Simulations do not provide proofs for theories, but they test their rational consistency and open novel windows to our understanding of complex phenomena. The present simulations show that to understand the evolution of diploid, genetically complex organisms, more sophisticated tools are required than those offered by simple linear or analytical models of haploid organism. For example, analytical approaches do not grasp the subtle complexities of aspects of inclusive fitness that explain actual biological evolution [54], or that are too remote from natural phenomena to be relevant for our understanding of biological evolution [55]. The simulation results presented here strongly suggest that synergy plays a central role in driving evolution, as was predicted by Hamilton [21] and shown by Queller [56-57] and Jaffe [22]. Evidently, a rational explanation for the evolution of sex must consider poliploidy, synergies that merge from reproductive division of labor and anisogamy, intergenerational effects of fitness and complexity. These can only be analyzed using more sophisticated tools than those developed by classical mathematics.

A special case is the modern vision about anisogamy. A universal pattern of sex roles may not exist [58]. Empirical data reveal an enormous variation in almost every aspect of sexual behavior and sex roles in a very broad range of animals. The results presented here suggest that if synergy is unleashed by the behavioral interactions among the sexes, evolution will favor a certain specific outcome. But, of course, many different outcomes are possible. And, of course, many different arrangements are possible.

A common criticism of complex simulations is that knowledge of the micro-macro-dynamics involved becomes fuzzy because of the excessive complexity involved. But robust trends often emerge. It is better to accept that our knowledge has limits due to complexity than to accept a false truth just because it is simple. Analytical mathematics used in theoretical biology has limitations for tackling complex problems. In the case of models based on haploids, for example, the simulations presented here suggest that they simply make extrapolations that are not applicable to the evolution of diploids, the most common genome in living organisms. Switching to algorithmic mathematics, such as ABM, is important in advancing our understanding of complex issues, such as the evolution of sex and of synergistic cooperation in general [22, 59]. More sophisticated models will elucidate more aspects of this complex dynamics with implications for the understanding biological and cultural evolution, intelligence, and complex systems in general.

# Acknowledgment.

I thank Adam Russell of DARPA for his enthusiastic promotion of a social-supercollider, first proposed by Duncan Watts, which influenced the organization of this paper, Guy Hoelzer for encouragement and for reminding me of Atmar's paper, Cristina Sainz and Zuleyma Tang-Martinez for helping improve the readability of the paper and to the late John Maynard Smith and William Hamilton for illuminating discussions. I profited from the constructive comments of several referees. Sonya Bahar did an excellent editorial work.

**Appendix.** Terms used in the academic literature that are used here:

| |
|---|
| **Anisogamy**: Condition in which reproductive gametes which fuse differ in chemistry, size, morphology and/or motility. |
| **Apomixis**: Asexual reproduction without fertilization (synonym to Parthenogenesis). |
| **Asexual**: Individual that produce offspring partenogenthetically or through thelytoky. |
| **Assortation**: Sorting or arranging according to characteristic or class. Self seeks like. |
| **Bisexual**: Two sexes are required for reproduction. |
| **Crossovers**: Two chromosomes break and then reconnect but to different end piece. Cromosomes are from different individuals in sexual reproduction and for the same individual among monosexuals. |
| **Epistatic**: Phenotypic expression of a gene is dependent on the presence of other genes. |
| **Gametogenesis**: Process in which cells undergo meiosis to form gametes. |
| **Hermaphrodite**: Individual displaying both male and female. |
| **Meiosis**: Cell division that reduces the number of chromosomes in the parent cell by half and produces four gamete cells. |
| **Mitosis**: Part of the cell cycle when replicated chromosomes are separated into two new nuclei. This includes DNA replication followed by an assignment of a full set of chromosomes into each of two new cells containing roughly equal shares of genes from each parent. |
| **Monosexual**: No sex but diploids suffer crossover between sets of chromosomes (Synonym to Thelytoky). Does not refer to monosexuality in plants. |
| **Mutation**: Random change that occurs in our DNA sequence. |
| **Parthenogenesis**: Reproduction without fertilization (synonym to apomixis). Reproduction by cloning. |
| **Ploidy**: Number of sets of chromosomes in a cell of an organism. **Haploid** means one set, **Diploid** two sets, and **Haplodiploid** one set in males and two in females. |
| **Thelytoky**: Females are produced from unfertilized eggs but diploids suffer crossover between sets of chromosomes (synonym to monosexual). |
| **Trisexual**: Three sexes are required for reproduction. See details in [28]. |